%% file: main.tex
\pdfoutput=1
\documentclass[12pt,a4paper]{article}

\usepackage{ifthen} 
\newboolean{pdflatex}
\setboolean{pdflatex}{true} 

\newboolean{articletitles}
\setboolean{articletitles}{true} 

\newboolean{uprightparticles}
\setboolean{uprightparticles}{false} 

\def\paperauthors{P. Billoir, T. Boettcher, M. De Cian, L. H. Uecker} 
\def\paperasciititle{Track fitting at the full LHC collision rate} 
\def\papertitle{Track fitting at the full LHC collision rate} 
\def\paperkeywords{{High Energy Physics}, {LHCb}} 
\def\paperlicence{CC BY 4.0 licence}
\def\paperlicenceurl{https://creativecommons.org/licenses/by/4.0/}

\newif\ifEnableSectionTOCLinks
\EnableSectionTOCLinksfalse 

\input{preamble}
\usepackage{longtable} 
\usepackage{subcaption}
\usepackage{dsfont}

\begin{document}

\renewcommand{\thefootnote}{\fnsymbol{footnote}}
\setcounter{footnote}{1}

\input{title-LHCb-PAPER}


\renewcommand{\thefootnote}{\arabic{footnote}}
\setcounter{footnote}{0}


\cleardoublepage


\pagestyle{plain} 
\setcounter{page}{1}
\pagenumbering{arabic}


\input{body}

\addcontentsline{toc}{section}{References}
\bibliographystyle{LHCb}
\bibliography{main,standard,LHCb-PAPER,LHCb-CONF,LHCb-DP,LHCb-TDR}

\newpage

\end{document}

%% file: preamble.tex
\usepackage[top=1in, bottom=1.25in, left=1in, right=1in]{geometry}

\usepackage{microtype}
\usepackage{lineno}  
\usepackage{xspace} 
\usepackage{caption} 

\usepackage{graphicx}  
\usepackage{color}
\usepackage{colortbl}
\graphicspath{{./figs/}} 

\usepackage{amsmath} 
\usepackage{amssymb}
\usepackage{amsfonts}
\usepackage{upgreek} 

\newcommand*\patchAmsMathEnvironmentForLineno[1]{%
\expandafter\let\csname old#1\expandafter\endcsname\csname #1\endcsname
\expandafter\let\csname oldend#1\expandafter\endcsname\csname
end#1\endcsname
 \renewenvironment{#1}%
   {\linenomath\csname old#1\endcsname}%
   {\csname oldend#1\endcsname\endlinenomath}%
}
\newcommand*\patchBothAmsMathEnvironmentsForLineno[1]{%
  \patchAmsMathEnvironmentForLineno{#1}%
  \patchAmsMathEnvironmentForLineno{#1*}%
}
\AtBeginDocument{%
\patchBothAmsMathEnvironmentsForLineno{equation}%
\patchBothAmsMathEnvironmentsForLineno{align}%
\patchBothAmsMathEnvironmentsForLineno{flalign}%
\patchBothAmsMathEnvironmentsForLineno{alignat}%
\patchBothAmsMathEnvironmentsForLineno{gather}%
\patchBothAmsMathEnvironmentsForLineno{multline}%
\patchBothAmsMathEnvironmentsForLineno{eqnarray}%
}

\usepackage[pdftex,
            pdfauthor={\paperauthors},
            pdftitle={\paperasciititle},
            pdfkeywords={\paperkeywords}]{hyperref}
\usepackage{hyperxmp}
\hypersetup{
    pdflicenseurl={\paperlicenceurl}
}

\usepackage[colorinlistoftodos,textsize=scriptsize]{todonotes}

\usepackage[bottom,flushmargin,hang,multiple]{footmisc}

\usepackage[all]{hypcap} 

\input{lhcb-symbols-def} 

\hypersetup{
  colorlinks   = true, 
  urlcolor     = blue, 
  linkcolor    = blue, 
  citecolor    = red   
}

\ifEnableSectionTOCLinks
    \usepackage[explicit]{titlesec} 
    
    \let\oldcontentsline\contentsline
    \renewcommand

    \titleformat{\section}{\normalfont\Large\bf}{\hyperlink{tocsection.\thesection}{{\thesection} \parbox[t]{\dimexpr\textwidth-1pc}{#1}}}{1pc}{}

    \titleformat{\subsection}{\normalfont\bf}{\hyperlink{tocsubsection.\thesubsection}{{\thesubsection} \parbox[t]{\dimexpr\textwidth-1pc}{#1}}}{1pc}{}

    \titleformat{name=\section,numberless}[display]{}{}{0pt}{\normalfont\Huge\bfseries #1}
\fi

\usepackage{cite} 
\usepackage{mciteplus}

%% file: lhcb-symbols-def.tex
\usepackage{xspace}
\usepackage{upgreek}

\def\lhcb   {\mbox{LHCb}\xspace}

\def\babar  {\mbox{BaBar}\xspace}

\def\lhc    {\mbox{LHC}\xspace}

\def\rich   {RICH\xspace}
\def\richone {RICH1\xspace}
\def\richtwo {RICH2\xspace}

\def\MagUp {\mbox{\em Mag\kern -0.05em Up}\xspace}

\def\hltone {HLT1\xspace}
\def\hlttwo {HLT2\xspace}

\ifthenelse{\boolean{uprightparticles}}%
{

 \def\Pmu         {\ensuremath{\upmu}\xspace}

 \def\Ppi         {\ensuremath{\uppi}\xspace}

 \def\Ppsi        {\ensuremath{\uppsi}\xspace}

 \def\PDelta      {\ensuremath{\Delta}\xspace}
 \def\PXi         {\ensuremath{\Xi}\xspace}
 \def\PLambda     {\ensuremath{\Lambda}\xspace}
 \def\PSigma      {\ensuremath{\Sigma}\xspace}
 \def\POmega      {\ensuremath{\Omega}\xspace}
 \def\PUpsilon    {\ensuremath{\Upsilon}\xspace}
 \let\oldPi\Pi
 \def\PPi         {\ensuremath{\oldPi}\xspace}

 \def\PB      {\ensuremath{\mathrm{B}}\xspace}
 \def\PD      {\ensuremath{\mathrm{D}}\xspace}
 \def\PJ      {\ensuremath{\mathrm{J}}\xspace}
 \def\PK      {\ensuremath{\mathrm{K}}\xspace}
 \def\Ps      {\ensuremath{\mathrm{s}}\xspace}

 \def\thebaroffset{0.0em}
}
{

 \def\Pmu         {\ensuremath{\mu}\xspace}

 \def\Ppi         {\ensuremath{\pi}\xspace}

 \def\Ppsi        {\ensuremath{\psi}\xspace}
 
 \mathchardef\PDelta="7101
 \mathchardef\PXi="7104
 \mathchardef\PLambda="7103
 \mathchardef\PSigma="7106
 \mathchardef\POmega="710A
 \mathchardef\PUpsilon="7107
 \mathchardef\PPi="7105
 \def\PB      {\ensuremath{B}\xspace}
 \def\PD      {\ensuremath{D}\xspace}
 \def\PJ      {\ensuremath{J}\xspace}
 \def\PK      {\ensuremath{K}\xspace}
 \def\Ps      {\ensuremath{s}\xspace}

 \def\thebaroffset{0.18em}
}
\newcommand{\offsetoverline}[2][\thebaroffset]{\kern #1\overline{\kern -#1 #2}}%

\makeatletter
\ifcase \@ptsize \relax
  \newcommand{\miniscule}{\@setfontsize\miniscule{4}{5}}
\or
  \newcommand{\miniscule}{\@setfontsize\miniscule{5}{6}}
\or
  \newcommand{\miniscule}{\@setfontsize\miniscule{5}{6}}
\fi
\makeatother

\DeclareRobustCommand{\optbar}[1]{\shortstack{{\miniscule (\rule[.5ex]{1.25em}{.18mm})}
  \\ [-.7ex] $#1$}}

\def\mup        {{\ensuremath{\Pmu^+}}\xspace}
\def\mun        {{\ensuremath{\Pmu^-}}\xspace} 

\def\mumu       {{\ensuremath{\Pmu^+\Pmu^-}}\xspace}

\def\squark    {{\ensuremath{\Ps}}\xspace}

\def\pion   {{\ensuremath{\Ppi}}\xspace}

\def\pip    {{\ensuremath{\pion^+}}\xspace}

\def\kaon    {{\ensuremath{\PK}}\xspace}

\def\KorKbar {\kern \thebaroffset\optbar{\kern -\thebaroffset \PK}{}\xspace}

\def\Kp      {{\ensuremath{\kaon^+}}\xspace}
\def\Km      {{\ensuremath{\kaon^-}}\xspace}

\def\D       {{\ensuremath{\PD}}\xspace}

\def\DorDbar {\kern \thebaroffset\optbar{\kern -\thebaroffset \PD}\xspace}
\def\Dz      {{\ensuremath{\D^0}}\xspace}

\def\Dp      {{\ensuremath{\D^+}}\xspace}
\def\Dm      {{\ensuremath{\D^-}}\xspace}

\def\DpDm    {\ensuremath{\Dp {\kern -0.16em \Dm}}\xspace}

\def\theDstarp{{\ensuremath{\D^{*}(2010)^{+}}}\xspace}

\def\B       {{\ensuremath{\PB}}\xspace}

\def\BorBbar {\kern \thebaroffset\optbar{\kern -\thebaroffset \PB}\xspace}

\def\Bd      {{\ensuremath{\B^0}}\xspace}

\def\BdorBdbar {\kern \thebaroffset\optbar{\kern -\thebaroffset \Bd}\xspace}

\def\Bs      {{\ensuremath{\B^0_\squark}}\xspace}

\def\BsorBsbar {\kern \thebaroffset\optbar{\kern -\thebaroffset \Bs}\xspace}

\def\jpsi     {{\ensuremath{{\PJ\mskip -3mu/\mskip -2mu\Ppsi}}}\xspace}

\def\Y#1S{\ensuremath{\PUpsilon{(#1S)}}\xspace}

\def\LorLbar     {\kern \thebaroffset\optbar{\kern -\thebaroffset \PLambda}\xspace}

\newcommand{\decay}[2]{\mbox{\ensuremath{#1\!\to #2}}\xspace}

\def\to                 {\ensuremath{\rightarrow}\xspace}

\def\AT#1     {\ensuremath{A_{\mathrm{T}}^{#1}}\xspace}           

\def\C#1      {\ensuremath{\mathcal{C}_{#1}}\xspace}                       
\def\Cp#1     {\ensuremath{\mathcal{C}_{#1}^{'}}\xspace}                    
\def\Ceff#1   {\ensuremath{\mathcal{C}_{#1}^{\mathrm{(eff)}}}\xspace}        
\def\Cpeff#1  {\ensuremath{\mathcal{C}_{#1}^{'\mathrm{(eff)}}}\xspace}       
\def\Ope#1    {\ensuremath{\mathcal{O}_{#1}}\xspace}                       
\def\Opep#1   {\ensuremath{\mathcal{O}_{#1}^{'}}\xspace}                    

\newcommand{\aunit}[1]{\ensuremath{\text{\,#1}}}

\newcommand{\tev}{\aunit{Te\kern -0.1em V}\xspace}
\newcommand{\gev}{\aunit{Ge\kern -0.1em V}\xspace}
\newcommand{\mev}{\aunit{Me\kern -0.1em V}\xspace}
\newcommand{\kev}{\aunit{ke\kern -0.1em V}\xspace}
\newcommand{\ev}{\aunit{e\kern -0.1em V}\xspace}

\newcommand{\mevc}{\ensuremath{\aunit{Me\kern -0.1em V\!/}c}\xspace}
\newcommand{\gevc}{\ensuremath{\aunit{Ge\kern -0.1em V\!/}c}\xspace}
\newcommand{\mevcc}{\ensuremath{\aunit{Me\kern -0.1em V\!/}c^2}\xspace}
\newcommand{\gevcc}{\ensuremath{\aunit{Ge\kern -0.1em V\!/}c^2}\xspace}

\def\cm   {\aunit{cm}\xspace}

\def\fb   {\ensuremath{\aunit{fb}}\xspace}
\def\invfb   {\ensuremath{\fb^{-1}}\xspace}

\def\sec  {\ensuremath{\aunit{s}}\xspace}

\def\mhz  {\ensuremath{\aunit{MHz}}\xspace}
\def\khz  {\ensuremath{\aunit{kHz}}\xspace}

\def\gsim{{~\raise.15em\hbox{$>$}\kern-.85em
          \lower.35em\hbox{$\sim$}~}\xspace}
\def\lsim{{~\raise.15em\hbox{$<$}\kern-.85em
          \lower.35em\hbox{$\sim$}~}\xspace}

\def\pt         {\ensuremath{p_{\mathrm{T}}}\xspace}

\def\evtgen     {\mbox{\textsc{EvtGen}}\xspace}

\def\geant      {\mbox{\textsc{Geant4}}\xspace}

\def\photos     {\mbox{\textsc{Photos}}\xspace}

\def\pythia     {\mbox{\textsc{Pythia}}\xspace}

\def\tbyps      {\aunit{TB/s}\xspace}

\def\tell1  {TELL1\xspace}
\def\ukl1   {UKL1\xspace}

\newcommand{\eg}{\mbox{\itshape e.g.}\xspace}
\newcommand{\ie}{\mbox{\itshape i.e.}\xspace}



%% file: title-LHCb-PAPER.tex
\begin{titlepage}
\pagenumbering{roman}


\vspace*{-1.5cm}
\centerline{\large EUROPEAN ORGANIZATION FOR NUCLEAR RESEARCH (CERN)}
\vspace*{1.5cm}
\noindent
\begin{tabular*}{\linewidth}{lc@{\extracolsep{\fill}}r@{\extracolsep{0pt}}}
\vspace*{-1.5cm}\mbox{\!\!\!\includegraphics[width=.14\textwidth]{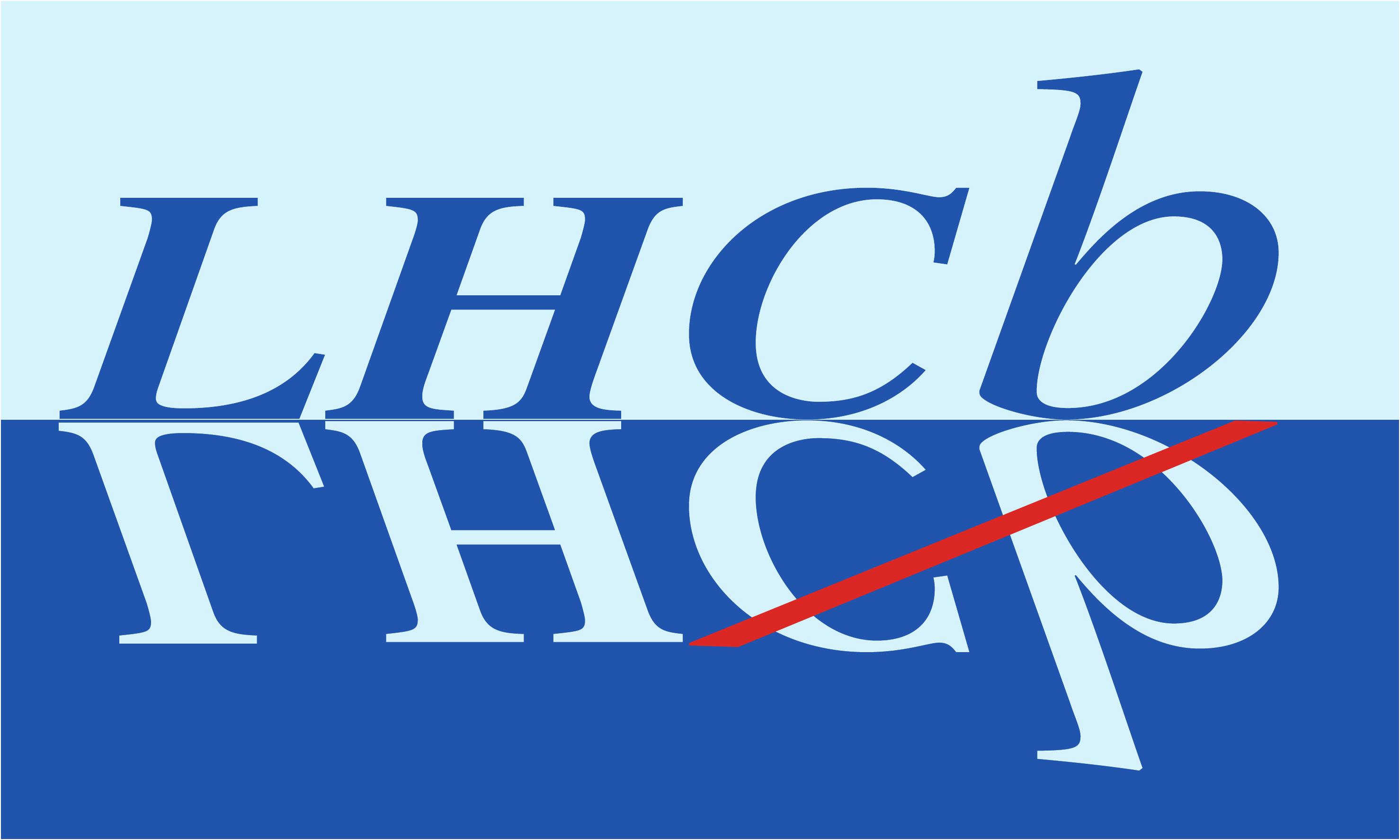}} & &
\\
 & & CERN-LHCb-RD-2026-003 \\  
 & & \today \\ 
 & & \\
\end{tabular*}

\vspace*{1.5cm}

{\normalfont\bfseries\boldmath\huge
\begin{center}
  \papertitle 
\end{center}
}

\vspace*{2.0cm}

\begin{center}
P. Billoir$^1$, T. Boettcher$^2$, M. De Cian$^3$, L. H. Uecker$^4$
\bigskip\\
{\normalfont\itshape\footnotesize
$^1$Laboratoire de Physique Nucl\'eaire et de Hautes Energies (LPNHE), Sorbonne Universit\'e,
CNRS/IN2P3, Paris, France \\
$^2$Department of Physics, Indiana University, Bloomington, USA \\
$^3$Department of Physics \& Astronomy, University of Manchester, United Kingdom\\
$^4$Physikalisches Institut, Ruprecht-Karls-Universit\"at Heidelberg, Heidelberg, Germany}
\end{center}


\begin{abstract}
  \noindent
  The \lhcb experiment at the Large Hadron Collider underwent a major upgrade before the LHC Run 3 data-taking period, employing an all-software approach in its trigger system. Here we present a fast implementation of a Kalman filter, used in the first trigger stage since the 2025 data-taking period, allowing to determine parameter estimates of charged-particle trajectories at the full \lhcb collision frequency of 30\mhz.
  This approach replaces computationally expensive magnetic field map lookups and numerical integration methods with fast analytical parameterisations while maintaining the mathematical framework of Kalman filtering. Implemented on approximately 500 GPUs within the first-level trigger, the algorithm has replaced the previous partial track fitting algorithm in the real-time trigger environment at the cost of a 2\% increase in processing time. Compared to the previous fitter this parameterised Kalman filter shows a significantly improved momentum resolution, resulting in a factor of two improvement in the invariant mass resolutions for reconstructed \Dz and \jpsi hadrons. It additionally demonstrates greater robustness against detector misalignment effects and substantially sharpens the discrimination between genuine particle trajectories and accidental background, more than doubling the rejection of the latter at no cost to genuine-track efficiency, for a standard selection working point.
    
\end{abstract}

\begin{center}
  Published in \href{https://link.springer.com/article/10.1140/epjc/s10052-026-16478-2}{Eur. Phys. J. C (2026) \textbf{86}: 1247}
\end{center}

\vspace{\fill}

{\footnotesize 
\centerline{\href{\paperlicenceurl}{\paperlicence}.}
}
\vspace*{2mm}

\end{titlepage}

%% file: body.tex
\newcommand{\parkf}{\texttt{Parameterised Kalman Filter}}
\newcommand{\vokf}{\texttt{VELO Only Kalman Filter}}
\def\Tesla {\aunit{T}\xspace}
\def\Tm {\aunit{Tm}\xspace}

\section{Introduction}
\label{sec:Introduction}

The \lhcb experiment \cite{LHCb-DP-2008-001, LHCb-DP-2022-002, LHCb-DP-2014-002} is one of the four big experiments at the Large Hadron Collider (LHC) at CERN. The original experiment and its upgrade are designed as general purpose detectors covering the forward direction. While LHCb's primary focus is heavy-flavour physics, its physics program also includes measurements from electroweak to heavy-ion physics, as well as a fixed-target program \cite{LHCb-TDR-020}.
 
The upgraded \lhcb detector \cite{LHCb-DP-2022-002} enables data collection at an instantaneous luminosity of $2 \times 10^{33}~\cm^{-2}\sec^{-1}$, a five-fold increase compared to the instantaneous luminosity during the previous data-taking period from 2015 to 2018.
For the upgrade, significant parts of the detector have been replaced or modified to be read out at the nominal LHC bunch-crossing frequency of 40\mhz: The tracking system has been fully replaced and consists of a silicon-pixel vertex detector (VELO) surrounding the pp interaction region\cite{LHCb-TDR-013}, a large-area silicon-strip detector (UT)\cite{LHCb-TDR-015}, located upstream of a dipole magnet with a bending power of about 4\Tm, and three stations of scintillating fibre detectors (SciFi) placed downstream of the magnet \cite{LHCb-TDR-015}. Charged hadrons are identified using information from two ring-imaging Cherenkov (\rich) detectors \cite{LHCb-DP-2012-003, LHCb-TDR-014}.
In order to fully profit from the increased instantaneous luminosity, a novel all-software trigger system is implemented to process, in its first stage, the full detector readout at the LHCb bunch-crossing frequency of 30\mhz\footnote{While the nominal \lhc bunch-crossing frequency is 40\mhz, the average proton-proton collision rate at \lhcb is about 30\mhz.} with a data rate of 4\tbyps.
This real-time strategy significantly increases the yield of hadronic and electron final states beyond the luminosity increase by circumventing the limitation of the previous hardware-based trigger system, which exploited calorimeter deposits for hadrons and relied on high-\pt signatures.
The trigger consists of two stages: \hltone is implemented on GPUs and performs a partial reconstruction with inclusive and exclusive kinematic and geometric selections based on approximately 50 different signatures, reducing the 30\mhz input rate to an output rate of about $1.2\mhz$ in 2025 \cite{LHCb-TDR-021,LHCb-DP-2022-002,Evans:2025jiw}. The events selected by \hltone are stored on a large disk buffer, where they are used to perform real-time detector alignment and calibration, before being processed by the second trigger stage, \hlttwo.
\hlttwo is implemented on a farm of CPUs and processes the buffered output of \hltone, performing the full offline-quality reconstruction. \hlttwo employs several thousand trigger selections to select a variety of physics processes, which are then either directly analysed, or passed on to a further offline selection step.
This approach has enabled \lhcb to record an integrated luminosity larger than 25\invfb at $\sqrt{s}=13.6\tev$ from 2024 to 2026, more than doubling the dataset collected during Run 1 and Run 2 of the LHC.

Central to the reconstruction in \hltone is the reconstruction of charged particle trajectories (tracks) and displaced decay vertices, which are key signatures for selecting events containing heavy-flavour hadron decays. \hltone is implemented using approximately 500 GPUs \cite{LHCb-DP-2021-003}, within the \texttt{Allen} \cite{Aaij:2019zbu} software project, achieving the required computational throughput for real-time processing at 30\mhz. 
Accurate determination of track parameters using a track fit is important for the \hltone selection performance. Up to the 2024 data-taking campaign, the \hltone track fitting algorithm performed a simplified fitting procedure by applying a Kalman filter to only the VELO segment of each track. This choice was originally driven by the need to meet the tight throughput constraints. 
Starting with the 2025 data-taking campaign, a new track fitting algorithm has been added to the reconstruction sequence of charged particles, the \parkf, performing a fit using information from all involved tracking detectors.

This paper is organised as follows: After a brief introduction of the \lhcb detector and its simulation, the principles of a Kalman Filter are reviewed in Sect.~\ref{sec:KF}, with the specifics in \lhcb discussed in Sect.~\ref{sec:parKF}. Section \ref{sec:Architecture} describes the implementation on GPUs, and Sect.~\ref{sec:Performance} analyses its performance in a simulated \lhcb detector. A conclusion is presented in Sect.~\ref{sec:Conclusion}.

\section{Detector and Simulation}
\label{sec:DetnSim}
\begin{figure}
    \centering
    \includegraphics[width=0.7\linewidth]{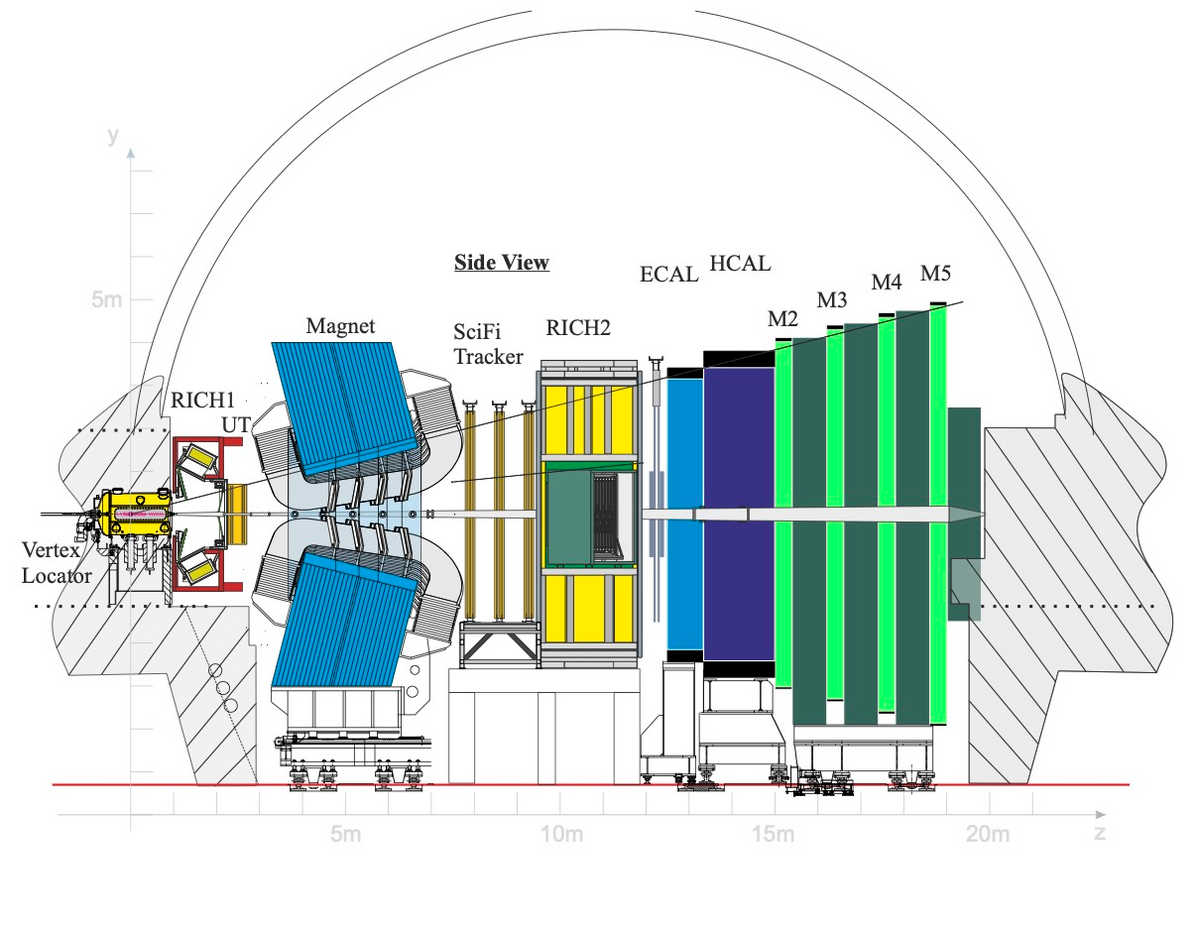}
    \caption{Layout of the upgraded LHCb detector. \cite{LHCb-DP-2022-002}}
    \label{fig:lhcb}
\end{figure}

The LHCb detector~\cite{LHCb-DP-2022-002} is a single-arm forward spectrometer covering the pseudorapidity range $2 < \eta < 5$, designed for the study of particles containing $b$ or $c$ quarks. A picture of its general layout can be seen in Fig.~\ref{fig:lhcb}. 
Its tracking system consists of the VELO with 52 stations, each covering half the azimuthal acceptance (in the $x-y$ plane). The stations are arrayed such that 99\% of all tracks from the region around the interaction point and within the nominal LHCb acceptance cross at least four VELO stations. The UT, a four layer silicon-strip detector, is situated downstream of the VELO in front of the large dipole magnet and is used to provide space point measurements before the volume with a large magnetic field. The SciFi is the main tracking detector downstream of the magnet. It consists of three stations with four layers each, two of them tilted at 5\textdegree\, to provide spatial information of the $y$ coordinate\footnote{The LHCb coordinate system is a right-handed cartesian system with the $z$ axis pointing downstream the detector along the beam line, and the $y$ axis pointing upwards.}. The dipole magnet creates a magnetic field, dominantly in the $y$ direction, with a peak field strength of about 1\Tesla and an integrated bending power of 4\Tm. In addition, \lhcb also operates two Ring Imaging Cherenkov (\rich) detectors, \richone and \richtwo. While \richtwo is located behind the SciFi tracker, \richone is located between VELO and UT and thus contributes significantly to the material budget to be considered in track fitting. The calorimetry systems and muon chambers, located behind \richtwo, are used to identify electrons, muons, photons and hadrons.

In the context of this work, simulation is used to determine the parameters of the \parkf\ and to evaluate its performance.
In the simulation, $pp$ collisions are generated using
\pythia~\cite{Sjostrand:2007gs,*Sjostrand:2006za}
with a specific \lhcb configuration~\cite{LHCb-PROC-2010-056}.
Decays of unstable particles
are described by \evtgen~\cite{Lange:2001uf}, in which final-state
radiation is generated using \photos~\cite{davidson2015photos}.
The interaction of the generated particles with the detector, and its response,
are implemented using the \geant
toolkit~\cite{Allison:2006ve, *Agostinelli:2002hh} as described in
Ref.~\cite{LHCb-PROC-2011-006}.
The studies presented here use simulated events of minimum bias pp-collisions, events containing a $\theDstarp \to (\Dz \to \Km \pip) \pip$ decay and events containing a $\Bs \to (\jpsi \to \mup \mun) (\phi \to \Kp \Km)$ decay, together representing a large part of the LHCb physics programme in terms of kinematics and topology.

\section{Track Fitting with a Kalman Filter}
\label{sec:KF}
Efficiently selecting collisions containing $b$ or $c$ hadron decays requires precise estimates of track parameters, including track momentum and direction.
Tracks are reconstructed in two steps: First the detector information is decoded and the resulting charged clusters in detectors with their spatial information, so-called hits, are combined to form particle trajectories using pattern recognition algorithms. Second, a track fit is performed to obtain the best estimate of the properties of the particle from the collection of associated hits. In \lhcb this fit is performed using a Kalman filter.

First published in 1960\cite{KalmanOriginal}, Kalman filters have now become a commonly used method in high energy physics. The Kalman filter's iterative nature has proven invaluable for reconstructing particle trajectories and reconstructing vertices from detector measurements, where the inherent measurement uncertainties and complex detector geometries present similar estimation challenges to those found in traditional non-particle physics applications\cite{FRUHWIRTH1987444, BILLOIR1984352}.
Kalman filters have also been used to fit the decay chains of heavy particles in the \babar \cite{Hulsbergen:2005pu} and \lhcb experiments, producing accurate uncertainties and correlations between quantities of interest.

Kalman filtering is an iterative process by which predictions from a physics model, here a charged particle moving through the detector and the magnetic field, are combined with measurement information from each detector layer to derive estimates of track parameters. The track information at a given position is called a state.
\lhcb is a forward detector, meaning all particles in the acceptance of the detector move in the positive $z$ direction. This simplifies the state description, as trajectories can be described as a function of the $z$ coordinate rather than the lifetime of the particle, and each detector element is roughly positioned at a fixed $z$ position.
For \lhcb a state is thus defined by the state vector $\vec{x}^\intercal = (x, y, t_x, t_y, \frac{q}{p})$ and its covariance matrix $P$ at given $z$ positions, usually at the position of a detector element.
The slopes $t_x = \frac{dx}{dz}$ and $t_y = \frac{dy}{dz}$ are given with respect to the $z$ axis and $\frac{q}{p}$ is the particle charge divided by the magnitude of its momentum, which ensures a near-gaussian distribution for this quantity\footnote{Due to the nature of measuring momenta via the bending of the trajectory in a magnetic field, $1/p$ and not $p$, with $p$ being the momentum, is (approximately) distributed according to a Gaussian distribution.}.

A Kalman filter needs to be initialised with an estimate of the starting state and its covariance. 
Filtering is then a repetition of prediction and update steps. 
In a prediction step, the state $\vec{x}_{k-1|k-1}$ and covariance $P_{k-1|k-1}$ in the layer $k-1$ are extrapolated to the layer $k$ using a function $\vec{f}_k$ and its Jacobian $F^{ij}_k = \frac{\partial f_k^i}{\partial x_k}$:
\begin{align}
    \vec{x}_{k|k-1} &= \vec{f}_k(\vec{x}_{k-1|k-1}), \\
    P_{k|k-1} &= F_k P_{k-1|k-1} F^{\intercal}_k + Q_k,
\end{align}
where the noise matrix $Q_k$ models the expected uncertainty due to multiple scattering of the charged particles in the material of the detector.
The predicted state $\vec{x}_{k|k-1}$ is then updated using the measurement $\vec{m}_k$ in the detector layer $k$ to obtain the updated state $\vec{x}_{k|k}$ and covariance $P_{k|k}$. 
First, a residual $\vec{r}_k$ is obtained by projecting the predicted state onto the measurement space with the matrix $H_k$, encoding the relevant geometry:
\begin{align}
    \vec{r}_k &= \vec{m}_k - H_k \vec{x}_{k|k-1}.
\end{align}
Then a covariance $S_k$ of the residual $\vec{r}_k$ is calculated by projecting the state uncertainty $P_{k|k-1}$ onto the measurement space and adding the measurement covariance $R_k$:
\begin{align}
    S_k &= H_k P_{k|k-1} H_k^\intercal + R_k.
\end{align}
The Kalman gain $K_k$ can be calculated from the state covariance $P_{k|k-1}$ and the covariance of the residual $S_k$ and describes how much the predicted state should be updated based on the residual:
\begin{align}
    K_k &= P_{k|k-1} H_k^\intercal S_k^{-1}.
\end{align}
From the Kalman gain $K_k$ and the residual vector $\vec{r}_{k}$ the updated values $\vec{x}_{k|k}$ and $P_{k|k}$ can be calculated:
\begin{align}
    \vec{x}_{k} &= \vec{x}_{k|k-1} + K_k \vec{r}_k, \\
    P_{k|k} &= (\mathds{1} - K_k H_k) P_{k|k-1}.
\end{align}
The $\chi^2$ value of an update step can be calculated as:
\begin{align}
    \chi^2_k &= \vec{r}_k^\intercal S_{k}^{-1}\vec{r}_k.
\end{align}
The reduced $\chi^2_{\mathrm{red}} = \sum_i^n\chi^2_i / \text{ndof}$, \ie the sum of all $\chi^2$ values divided by the degrees of freedom (ndof) of the track fit, can be used as a measure of the goodness-of-fit.

Using this procedure only the last state has the best parameter estimates. Often, however, Kalman filters are used as smoothers, which enables the algorithm to provide state estimates also incorporating information downstream of the position of interest. If it is of interest to achieve a precise state measurement $\vec{x}_{k}$, where $k$ is smaller than the total number of hits, $n$, an additional backwards pass can be performed. Starting at layer $n$ going down to zero, the propagation and filtering process is repeated. If the state of interest and its covariance is saved both in forwards and backwards pass, they can then be combined to a smooth state $\vec{x}_{k|n}$, incorporating the information of all measurements. In real applications it can be necessary to repeat forward and backward passes for numerical stability.

\section{The \parkf}
\label{sec:parKF}

Owing to the parallel nature of track fitting and the parallelisation capabilities of modern computing architectures, several studies have been performed with Kalman filter-based track fitting, see \eg Refs.~\cite{Ai2021,ParallelKalman1}. The implementation discussed in the following however is the first full Kalman filter that was used in a production environment of an LHC experiment at the full collision rate.

The \parkf\ is a Kalman filter as defined in the previous section, but uses simplifications specific to typical implementations in a high-energy physics context which allows it to be executed with minimal time consumption. It was developed as a drop-in replacement for the previous track fitting algorithm, the \vokf, to provide better momentum (and therefore mass) resolution, more robustness against detector misalignments, and improved rejection against fake tracks (so-called ghosts) in the \hltone stage of \lhcb.
The \vokf\, performs a simplified Kalman filter track fit of only the track segment in the VELO detector under the assumption of a vanishing magnetic field, full decoupling of the $x$ and $y$ components and a parameterised scattering description. These simplifications were originally made to stay within the given throughput constraints.
The \parkf\ extends the track fitting to all three tracking detectors: the VELO, UT and SciFi. First described as a fast alternative to a full Kalman filter \cite{LHCb-DP-2021-001}, and prototyped on CPUs, this approach has proven to be well suited for the GPU-based \hltone and provides improved state estimates while still fulfilling the real-time requirements.

\begin{figure}
    \centering
    \includegraphics[width=1.0\linewidth]{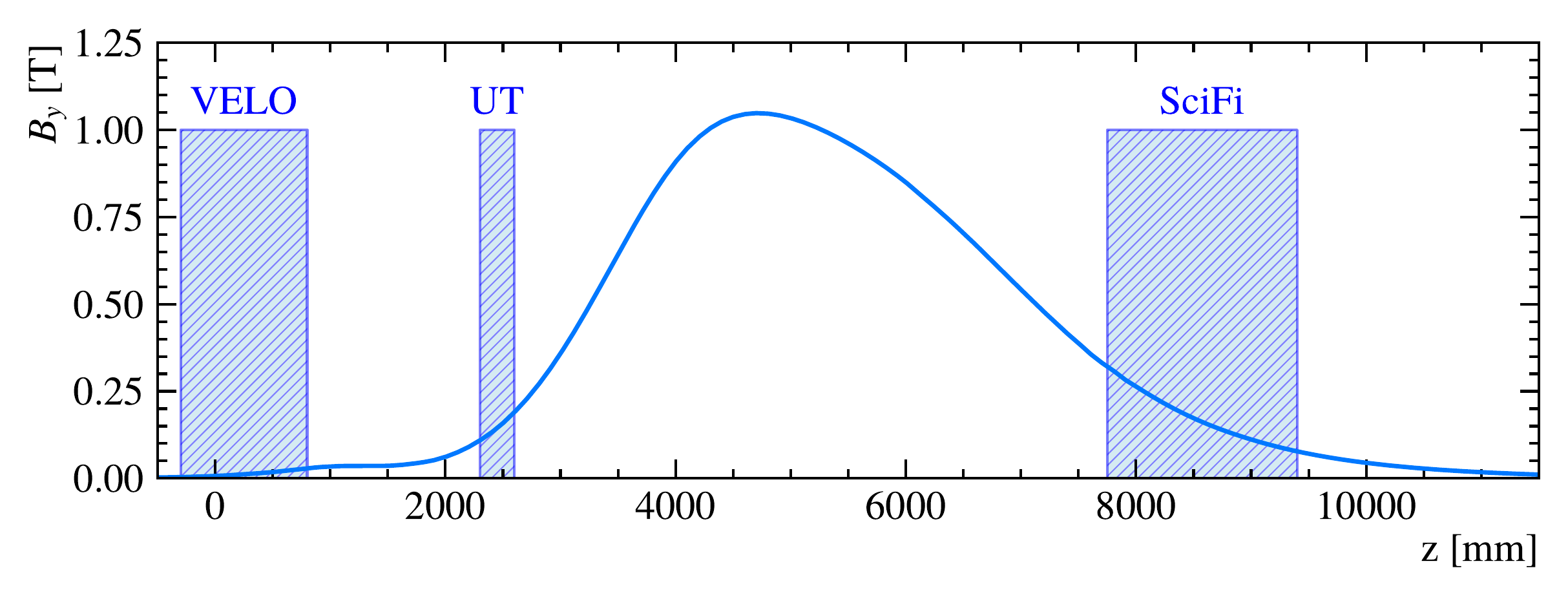}
    \caption{Measured magnetic field strength in the $y$ direction inside the \lhcb detector as a function of $z$\cite{LHCbBFieldMap}. Overlaid are the positions of the three tracking detectors.}
    \label{fig:magfieldmap}
\end{figure}

As a Kalman filter performs extrapolations of particle trajectories between layers, the standard implementation of such an algorithm includes frequent look-ups of a magnetic field map and the detector geometry, as well as intensive calculations to predict the movement of a charged particle in the magnetic field by numerically solving the equations of motion. The cost of such an implementation is substantial, in the CPU-based \hlttwo reconstruction, where a full Kalman filter is employed, the track fit accounts for approximately 14\% of the total processing time. This approach is not suited for \hltone as the field map and geometry exceed the GPU's limited fast memory, while the position-dependent lookups and data-dependent integration steps produce irregular, non-coalesced memory accesses and control-flow divergence across threads.
Almost all throughput improvements in the \parkf, compared to a standard implementation as described above, are achieved in changes to the state prediction step. Instead of extrapolating the state using numerical methods to solve the propagation of a charged particle in a magnetic field, a simple function is chosen to parameterise this extrapolation. The same is true for estimating multiple scattering contributions to the covariance matrix of the state, where simulation-derived parameterisations for each extrapolation steps are used. Different parameterisations are employed for each subdetector and for the steps between the subdetectors, with different sets of parameters for each layer, where the different propagation parameterisations are mainly driven by the strength of the magnetic field in the respective regions, see Fig.~\ref{fig:magfieldmap}. 
While the full parameterisation is explained in Ref. \cite{LHCb-DP-2021-001} and this implementation is largely using the same formulae, one instructive example is given here.

The propagation of state $\vec{x}_{k-1|k-1}$ in the layer $k-1$ to the layer $k$ is given by the empirical vector-valued function $\vec{f}_{X\to Y}$, where $X$ and $Y$ denote the specific subdetectors of the involved layer, \ie $\vec{x}_{k|k-1} = \vec{f}_{X\to Y}(\vec{x}_{k-1|k-1}; p^{X\to Y, k-1})$\footnote{The superscripts on the parameters will be dropped from here on for better readability.}. For example, the effect of the magnetic field is already sizeable between the UT layers, with a field strength in the $y$~component $B_y$ between 0.1\Tesla and 0.3\Tesla. In this region the extrapolation function takes the form:
\begin{equation}
    \vec{f}_{UT \to UT}(\vec{x}, \Delta z)=\left(\begin{array}{c}
        x+\left[p_{3} t_{x}+\left(1-p_{3}\right) f^{t_{x}}(\vec{x})\right] \Delta z \\
        y+\left[p_{5} t_{y}+\left(1-p_{5}\right) f^{t_{y}}(\vec{x})\right] \Delta z \\
        t_{x}+\left[p_{0} \frac{q}{p}+p_{1}\left(\frac{q}{p}\right)^{3}+p_{2} y^{2} \frac{q}{p}\right] \Delta z \\
        t_{y}+p_{4} \frac{q}{p} t_{x} \text{sign}(y) \\
        \frac{q}{p}
    \end{array}\right)
\end{equation}

Note that the $x/y$ components rely on the predicted slopes, here written with the shorthand $f^{t_{x/y}}(\vec{x})$ and $\Delta z$ is the extrapolation distance along the $z$ axis, which is calculated based on the detector geometry.

The parameterisation of noise resulting from multiple scattering is the same for all extrapolation steps. The underlying assumption is that the scattering angle $\theta$ follows a central normal distribution $\mathcal{N}$ centred around 0 with a width inversely proportional to the momentum, \ie $\theta \sim \mathcal{N}(0, \frac{\alpha}{p})$, where $\alpha$ is the parameter to be extracted from simulation. By extending this idea to two correlated dimensions and propagating the scattering angle to the state observable, the symmetric noise matrix $Q$ can be derived:
\begin{equation}
\begin{array}{cc}
     \sigma_x = \tilde{p}_{0} \left|\Delta z \frac{q}{p}\right|, & \sigma_{tx} = \tilde{p}_{2} \left|\frac{q}{p}\right|,\\
     \sigma_y = \tilde{p}_{1} \left|\Delta z \frac{q}{p}\right|, & \sigma_{ty} = \tilde{p}_{3} \left|\frac{q}{p}\right|,
\end{array} \\
\end{equation}
\begin{align}
\label{eq:Q}
\boldsymbol{Q}&=\begin{pmatrix}
\sigma_x^2 & 0 & \tilde{p}_{4} \sigma_x \sigma_{t_x} & 0 & 0 \\
0 & \sigma_y^2 & 0 & \tilde{p}_{5} \sigma_y \sigma_{t_y} & 0 \\
\tilde{p}_{4} \sigma_x \sigma_{t_x} & 0 & \sigma_{t_x}^2 & 0 & 0 \\
0 & \tilde{p}_{5} \sigma_y \sigma_{t_y} & 0 & \sigma_{t_y}^2 & 0 \\
0 & 0 & 0 & 0 & 0
\end{pmatrix}.
\end{align}
This includes four parameters for the scaling of the scattering with momentum, $\tilde{p}_{0}, \tilde{p}_{1}, \tilde{p}_{2}, \tilde{p}_{3}$, and two parameters for the correlation between slope and position, $\tilde{p}_{4}, \tilde{p}_{5}$. For the scattering in the VELO, the parameters for $x$ and $y$ are assumed to be equal, halving the size of the parameter set. This is a reflection of the weak magnetic field in the extrapolation making a symmetric approximation possible. Outside the VELO this assumption no longer holds, as the magnetic field singles out the $x$ plane and the scattering is defined including the extrapolation. The fitted parameters for $x$ and $y$ differ by around 10\%, showing that the difference cannot be neglected. While multiple scattering is taken into account as such, energy loss is not directly accounted for, \ie $f_{X\to Y}^{\frac{q}{p}}(\frac{q}{p}) = \frac{q}{p}$. The ratio $\frac{q}{p}$ is thus defined as the value at the production of the particle and the energy loss is absorbed into the parameterisation of the propagation of the remaining state components $(x, y, t_x, t_y)$ as well as the scattering matrices $Q$. The validity of this approach can be seen in the unbiased momentum estimates presented in Sect.~\ref{sec:Performance}.
To propagate the covariance matrix $P$ alongside the state from one detector position to another, the Jacobians of the extrapolation functions need to be calculated and are given by $F_{ij} = \frac{\partial f_i}{\partial x_j}$.
These matrix elements can be calculated analytically as derivatives of the extrapolation functions.

For the \parkf\ to work correctly, the extrapolation functions need to be precise enough such that the parameterisation uncertainty $\delta_{\text{par}} = x_{\text{true extrap}} - x_{\text{param. extrap}}$ is smaller than the physical uncertainties of the state. There are two sources of state uncertainty included in the \parkf: the contribution from multiple scattering, $Q$, and the inherent resolution of the tracking detectors, \eg from the non-zero pixel size. 
For all particles that did not experience any hard momentum changes, \eg bremsstrahlung emission, multiple scattering (MS) is the dominant source of uncertainty on the state estimates, with limited detector resolution becoming significant for particles with momenta in excess of 100\gevc.
The parameterisations therefore need to be accurate to a precision better than the MS contribution. As the asymptotic case for tracks of infinite momentum, \ie $\frac{q}{p} \to 0$, is no bending in the magnetic field, the parametrisation functions need to recover a linear function in that limit to maintain an asymptotically correct $\delta_{\text{par}}$ distribution.
Energy loss not described by MS, especially the emission of bremsstrahlung by electrons, is not accounted for in the fit model. The recovery of bremsstrahlungs emission is performed by a separate algorithm based on calorimeter information. The selection algorithms involving electrons are designed to take their energy-loss characteristics into account.

In the event processing of \hltone, only a single track state is used. This is the `closest-to-beamline' state, \ie the point of closest approach of the track extrapolation to the LHC beamline. This state is used in single-track trigger selections and to reconstruct secondary vertices. Saving only a single state enables further simplification of the usual Kalman filter process, which combines forward and backward passes and averages the intermediate states for the best possible estimates. The minimal configuration chosen is a full forward pass and a backward pass only over the VELO detector.
For the partial backwards pass, the covariance matrix at the end of the forward pass is transported back to the end of the VELO using a cumulative transport matrix, then the backwards pass of the VELO is performed starting with the forwards pass state at that position with an updated momentum estimate. 
An accurate momentum estimate is achieved from the forward pass alone and the `closest-to-beamline' state is almost completely determined by the VELO backward pass. Given this configuration, no outlier removal, \ie the removal of hits above a threshold contribution to the track $\chi^2$, is performed. The output of the algorithm is reduced to the position ($x$, $y$, $z$), slopes ($t_x$, $t_y$), charge over momentum $\frac{q}{p}$, a simplified covariance ($P_{xx}$, $P_{xt_{x}}$, $P_{t_{x} t_{x}}$, $P_{yy}$, $P_{yt_{y}}$, $P_{t_{y} t_{y}}$) and track fit quality information ($\chi^2$, ndof). This is all the information needed for an accurate determination of the four-momenta of the particles (assuming a mass) and the positions and uncertainties of the primary and secondary vertices, and their related quantities such as impact parameter significance.

\section{Implementation}
\label{sec:Architecture}
GPU architectures achieve high throughput with massive parallelism rather than low-latency operations. Computation is expressed in kernels, functions executed on the GPU by a large number of threads working in parallel. Threads are grouped into blocks of threads that can cooperate, and further divided into a warps, the basic unit of parallelisation, which perform operations in lockstep using 32 simple compute units. The quantity and quality of the basic compute units is dependent on the architecture: on the chipset used by HLT1 (NVIDIA RTX A5000) single precision floating point operations are preferred. When different threads within a warp need to perform different operations, the execution is serialised and the throughput drops. Thus, algorithms with uniform and data-independent execution paths are preferred. Memory access is an important bottleneck as there is a hierarchy of memory spaces with increasing latency. The registers file is the fastest available and directly connected to the compute units, the global off-chip memory can store large amounts of data, but accesses are associated with large latency. Additional on-chip caches with intermediate latencies exist. Latency is not only hidden by caches, but also by switching between warps to hide the latency of memory access.

The independence of each Kalman filter track fit from any other makes it easy to utilise the parallelism of the GPU architecture. At the same time, each individual track cannot easily be parallelised on a per-hit level since each step of the Kalman filter depends on the result of the previous step. Based on these two considerations the choice is made to have each individual thread perform one track fit at a time. This enables having tracks from different events processed in the same kernel block to fully utilise the available threads. As a direct consequence the runtime of the algorithm scales linearly with the number of tracks in an event. Further, it is necessary to consider the numerical precision limitations. The double floating point precision, commonly used in CPU applications, would be penalised with a performance factor of 1:64 on the specific chipset on which \hltone is implemented in Run 3. The \parkf\ performs all calculation in single floating point precision by default in order to maximize throughput. This makes full use of the available processor cores optimised for single precision and reduces the memory resources allocated to a single track and thus increases opportunities for parallelisation. The reduction in numerical precision leads to small differences in the parameter estimates compared to the same algorithm running in double floating point precision, however the impact on the parameter estimates is acceptable for HLT1.

For the processing of each track a large amount of data is needed, specifically the number of floating point operations per byte of memory access is low, this means the kernel is limited by the available memory resources. Each thread needs to access the (unique) hit information and the common detector information. This adds up to several kilobytes per thread. The use of parameterisations, rather than numerically calculating the extrapolation, solves this issue by replacing unique access to a magnetic field map with a uniform access to parameters, \eg the extrapolation of a track within the vertex detector is sufficiently described by six parameters regardless of the track position. These accesses can further be optimised by using the constant memory of the GPU. Constant memory is a 64~kB large read-only memory space with a broadcast ability that allows the distribution of data to all threads in a warp without additional latency. For the extrapolation between hits, each thread uses the same parameters in lockstep and the size of the parameterisation dataset is around 2~kB, so it can be optimised with the constant memory. The exception to this access pattern is the extrapolation through the main dipole magnet, where each thread needs to access a different subset of the much larger parameterisation dataset around 1~MB. This dataset is too large to be stored in the constant memory and the access pattern would not profit from the broadcasting capabilities. Instead the parameterisation dataset is stored in the global memory space, and in practice cached in on-chip memory. By designing the algorithm to interweave memory accesses with compute operations the latency of the memory can be partially hidden.

In general, the algorithm is designed to give as much opportunity as possible for the kernel to hide latency. Reducing the amount of data that is stored in the registers has led to major throughput improvements. When the algorithm needs more space than there is register space, register spilling occurs, this is the slow offloading of register-stored data into slower memory. The maximum register space used by a thread also determines the occupancy, \ie, how many warps can share the same compute units, and thus how effectively the GPU can hide latency. The goal is to keep track parameters, such as the state $\vec{x}$ and its covariance in the registers, while all other data structure are broken up into small objects only to be kept in fast memory for short periods.
The data structures of external inputs, such as the track objects and the detector hits, are given by upstream algorithms and they are not necessarily optimal for the access pattern of the Kalman filter. For example, in the case of the SciFi and UT track segments, the order of hits, as stored in the track object, does not match the processing order in the Kalman filter. In these cases lightweight maps of hits to layers are created to avoid unnecessary reading processes. Since each track has the same structure, branching instructions can mostly be avoided and the loop structure over all hits in $z$ order can be synchronised on a warp level.

The parameters used in the extrapolation are derived from simulated events. For each extrapolation step a dataset $D$ is created, which contains the true state at the origin of the extrapolation step, $\vec{x}_{n,\text{origin}}$, the true state at the target of the extrapolation, $\vec{x}_{\text{target}}$, and a state extrapolated from the origin state to the target $z$ layer using a precise Runge-Kutta extrapolation, $\vec{x}_{n,\text{extrap}}$. The important difference between the latter two is that $\vec{x}_{n,\text{target}}$ contains a contribution of multiple scattering. The extrapolation parameters $p_i^{X\to Y}$ are obtained by minimising 
\begin{equation}
    \prod_{n\in D} \mathcal{N} \left( f^{a}_{X \to Y}(\vec{x}_{n, \text{origin}}; p) - a_{n, \text{extrap}}, \theta \right),
\end{equation}
where $a \in \{ x, y, t_x, t_y\}$ is the state component.
This effectively ensures the parameterised extrapolation matches the previously calculated Runge-Kutta extrapolation. The width $\theta$ of the Gaussian used in the minimisation describes the parameterisation uncertainty. This minimisation is done for the individual dimension $a$ of the state in the order that respects the dependencies in the extrapolation function, \eg the predicted $x$ value depends on the predicted $t_x$ value. The scattering parameters $\tilde{p}$ in $x$ are extracted by fixing the extrapolation parameters $p$ and minimising with only the scattering parameters $\tilde{p}$ left floating:
\begin{equation}
  \prod_{n\in D} \big [ \mathcal{N}_2 \bigl(
    f^{a}_{X \to Y}(\vec{x}_{n,\text{origin}};\, \tilde{p})
      - a_{n,\text{target}},
    f^{t_a}_{X \to Y}(\vec{x}_{n,\text{origin}};\, \tilde{p})
      - t_{a|n,\text{target}},
    Q_{[at_a]} \bigr) + c \big ],
\end{equation}
where $a \in \{x,y \}$ denotes the spatial dimension, $Q_{[at_a]}$ is the relevant $2 \times 2$ sub matrix of $Q$ (Eq.~\ref{eq:Q}) and $c$ is an empirical constant to help with convergence. Relatively small sample sizes of 10,000 simulated events lead to 100,000 - 1,000,000 data points in $D$, depending on the specific extrapolation function and are sufficient to determine all parameters. The required accuracy, better than the expected MS contribution, is achieved for all extrapolation functions, \ie the mean squared deviation of the parameterised extrapolation is less than the expected uncertainty resulting from MS.
Additionally, the parameterisation is extended to be valid for momenta as low as 1.5\gevc, rather than 3\gevc as in the first report on this algorithm\cite{LHCb-DP-2021-001}, matching the output of the \hltone track reconstruction algorithms.
As the \parkf\ uses the same set of parameters for both polarity settings of the main \lhcb dipole magnet by applying a sign flip to extrapolation terms sensitive to the sign of $\frac{q}{p}$, the parameters fit also uses a dataset containing events of both magnet polarities. The parameter set comprises 240,466 floating-point values, of which 240,000 describe the extrapolation from the UT to the SciFi, given by a grid of size $50 \times 60$, and each grid point associated with an 80-parameters polynomial. Each thread reads 946 of them: the 480 extrapolation parameters relevant to its track's region via interpolation of 6 grid points in the UT to SciFi extrapolation\cite{LHCb-DP-2021-001}, together with the remaining 466 values, which are common to all threads. These 466 shared values are placed in constant memory, where their broadcast access across threads is served efficiently.

The \parkf\ algorithm makes up roughly 3.5\% of the total runtime of the 2025 \hltone sequence, while the \vokf\ made up 1.4\% of the runtime, in an otherwise equal sequence.
This modest runtime increase from the \parkf\ does not impact the overall system throughput. While the evaluation was performed at the nominal pile-up of $\nu=7.6$, it is important to note that the throughput of the algorithm scales linearly with pile-up.
\lhcb expects an input rate to \hltone of about 30\mhz, dictated by the collision rate delivered by the \lhc to \lhcb. With 500 processors working in parallel, the minimum viable throughput for a single GPU is about 60\khz.
However, some safety margin needs to be accounted for to deal with outages, so any viable configuration must achieve throughputs well over 60\khz per GPU. 

The impact of the \parkf\ in terms of throughput is shown in Table~\ref{tab:throughput}. From all these numbers it is clear that the \parkf\ is well compatible with \hltone's real-time requirements. 

\begin{table}
\caption{Comparison of throughput on a single production GPU (NVIDIA RTX A5000) running the default HLT1 sequence for 2024 with the two different track fits algorithms. The results and their uncertainty are extracted from 10 individual runs of the same setup, emulating the production environment and their variance.}
\label{tab:throughput} \centering      
    \begin{tabular}{ll}
    \hline\noalign{\smallskip}
    Sequence & Throughput (events/s/device) \\
    \noalign{\smallskip}\hline\noalign{\smallskip}
    VELO-only KF & $65\,940 \pm 160$ \\
    Parameterised KF & $64\,310 \pm 90$ \\
    \noalign{\smallskip}\hline
    \end{tabular}
\end{table}

\section{Performance Analysis}
\label{sec:Performance}

\begin{figure}
\centering
\begin{subfigure}{.45\textwidth}
  \centering
  \includegraphics[width=.95\linewidth]{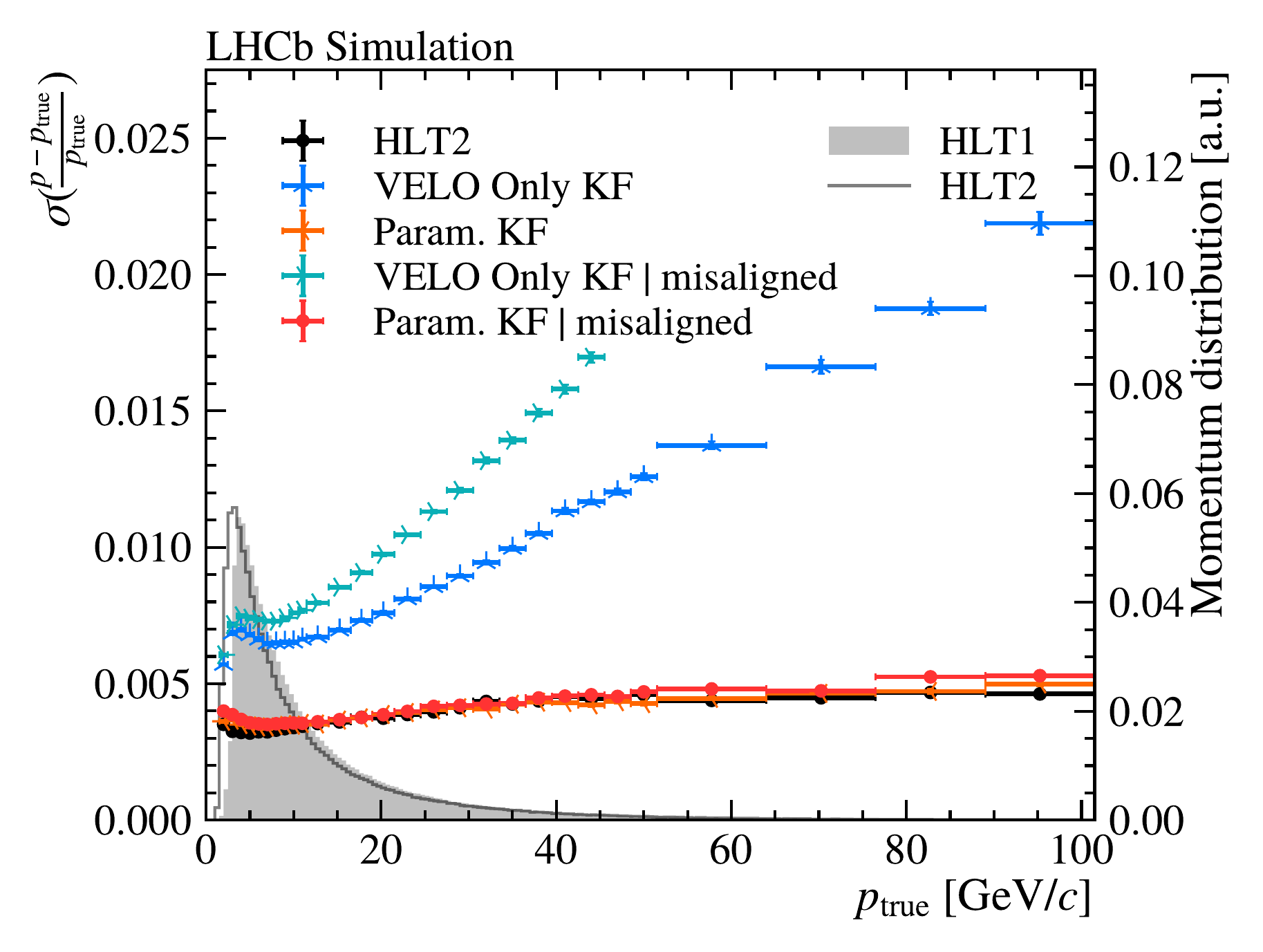}
  \caption{Differential momentum resolution as a function of the true momentum.}
  \label{fig:p_res_all}
\end{subfigure}%
\hspace{0.05\textwidth}
\begin{subfigure}{.45\textwidth}
  \centering
  \includegraphics[width=0.95\linewidth]{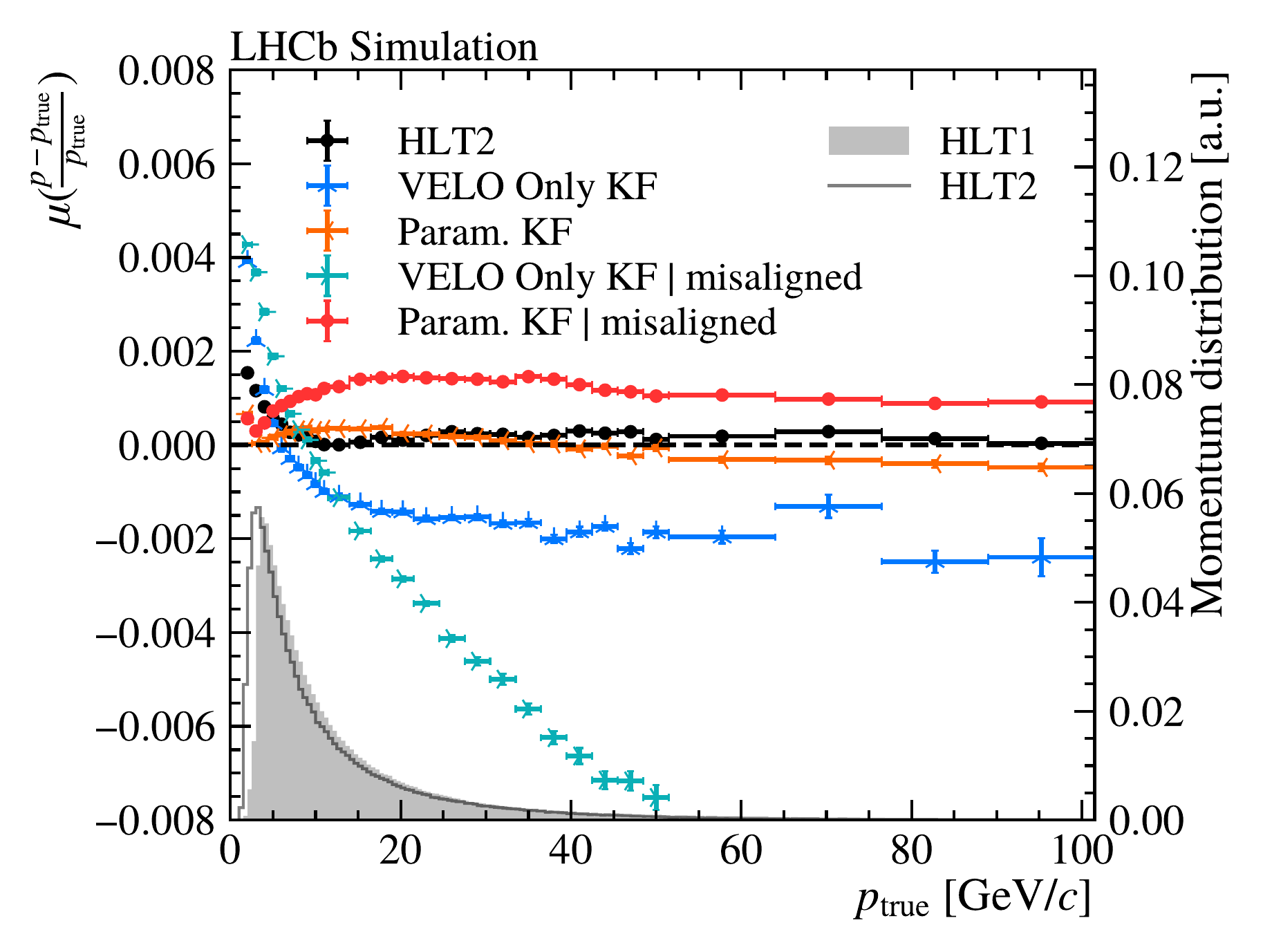}
  \caption{Differential momentum bias as a function of the true momentum.}
  \label{fig:p_bias_all}
\end{subfigure}
\caption{The momentum resolution is compared using the same simulated events of the \parkf\ and the algorithm it replaces the \vokf, both with and without misalignment injected into the simulated geometry of the sample. The momentum resolution of \hlttwo is included as a comparison point. The momentum resolution is extracted from a Gaussian fit to $(p - p_{\text{true}}) / p_{\text{true}}$ in the given bin of $p_{\text{true}}$, the ground truth value of the simulation.}
\label{fig:p_all}
\end{figure}

The primary benefit of fitting the full track comes from incorporating hits upstream and downstream of the \lhcb magnet, which enables precise measurements of particle momenta. The resulting momentum resolution is much better than that achieved by the pattern recognition algorithms (using a parametric approach) and used by the \vokf. The improvement in momentum resolution is shown as a function of momentum in Fig.~\ref{fig:p_res_all}, achieving a momentum resolution close to the one of the full reconstruction in \hlttwo \cite{LHCb-DP-2022-002}. In addition the figure also displays the impact of detector misalignment on the momentum estimate. For this study the detector elements of the SciFi detector only have been shifted randomly in the simulation, with the size of the shifts being realistic for the 2024 data-taking period. One can see the large impact on the resolution for the \vokf, while the \parkf\ manages to fully recover the performance of a perfectly aligned detector. In Fig.~\ref{fig:p_bias_all} a similar effect can be seen for the mean bias of the momentum estimate. As both algorithm treat the fit of the VELO segment identically, there is no difference in sensitivity to the VELO alignment.

\begin{figure}
\centering
\begin{subfigure}[t]{.45\textwidth}
  \centering
  \includegraphics[width=.95\linewidth]{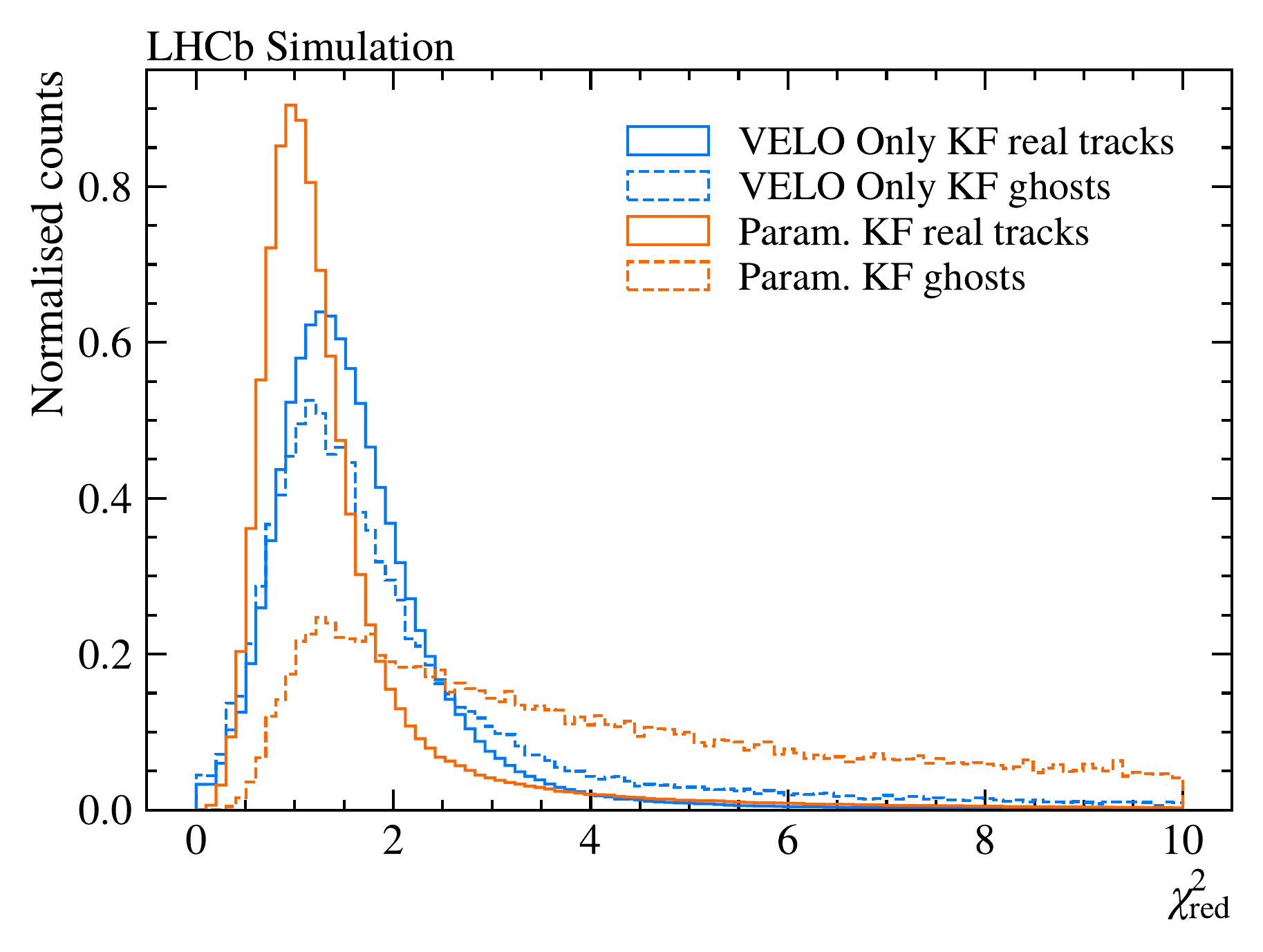}
  \caption{Distributions of the $\chi^2_{\mathrm{red}}$ values.}
  \label{fig:chi2_dist_small}
\end{subfigure}%
\hspace{0.05\textwidth}
\begin{subfigure}[t]{.45\textwidth}
  \centering
  \includegraphics[width=0.95\linewidth]{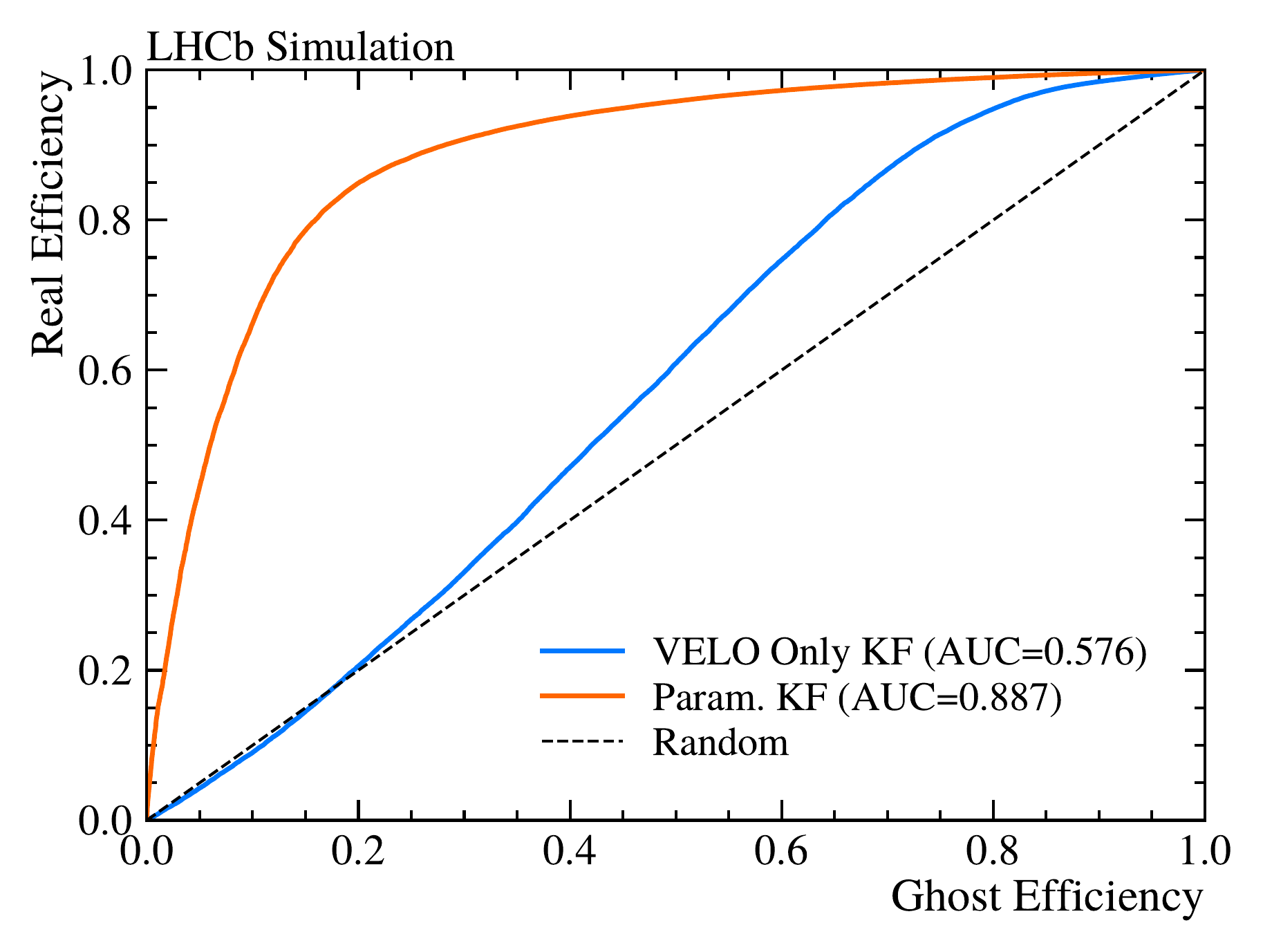}
  \caption{Receiver Operating Characteristic (ROC) curve of the ghost rejection.}
  \label{fig:chi2_roc}
\end{subfigure}
\caption{Discrimination power between real and fake (ghost) tracks, compared using the same simulated events, with the \parkf\ and the algorithm it replaces, the \vokf. Shown are (left) the $\chi^2_{\mathrm{red}}$ distributions and (right) the ghost-rejection power represented as an ROC curve. A ghost track is defined as a track for which less than 70\% of its hits originate from a single simulated particle.}
\label{fig:chi2_dist}
\end{figure}

The $\chi^2$ value produced by the Kalman filter is also used to select tracks of high quality or reject tracks that are likely the combination of random hits, leading to ghost tracks. A neural-network based ghost rejection is already used as part of the track reconstruction, resulting in an average fraction of ghost tracks considered in the track fit of only around 3\%. Further reduction of this fraction is performed with a cut on the quality-of-fit measure, $\chi^2_{\text{red}}$.
The $\chi^2_{\text{red}}$ calculated by the \vokf\ only accounts for the VELO segments, compared to the \parkf\ which accounts for the full track. Many ghost tracks are combinations of real VELO segments with real SciFi segments from different particles. This type of ghost track cannot be identified by a fit of only the VELO segment, thus the \parkf\ creates a larger separation between the $\chi^2_{\text{red}}$ distributions of real tracks and ghosts compared to the \vokf, as seen in Fig.~\ref{fig:chi2_dist}. Especially selections of low $\chi^2_{\text{red}}$ values result in a higher purity of the track sample. At the working points used by the high-rate single track selection, the VELO-only fit ($\chi^2_{\text{red}} < 2.5$) produces a ghost efficiency of 70\% at a genuine track efficiency of 87\%, while the \parkf\ ($\chi^2_{\text{red}} < 3.0$) improves to a ghost efficiency of 29\% with the efficiency for genuine tracks simultaneously increasing to 91\%.

\begin{figure}
\centering
\begin{subfigure}[t]{.45\textwidth}
  \centering
  \includegraphics[width=.95\linewidth]{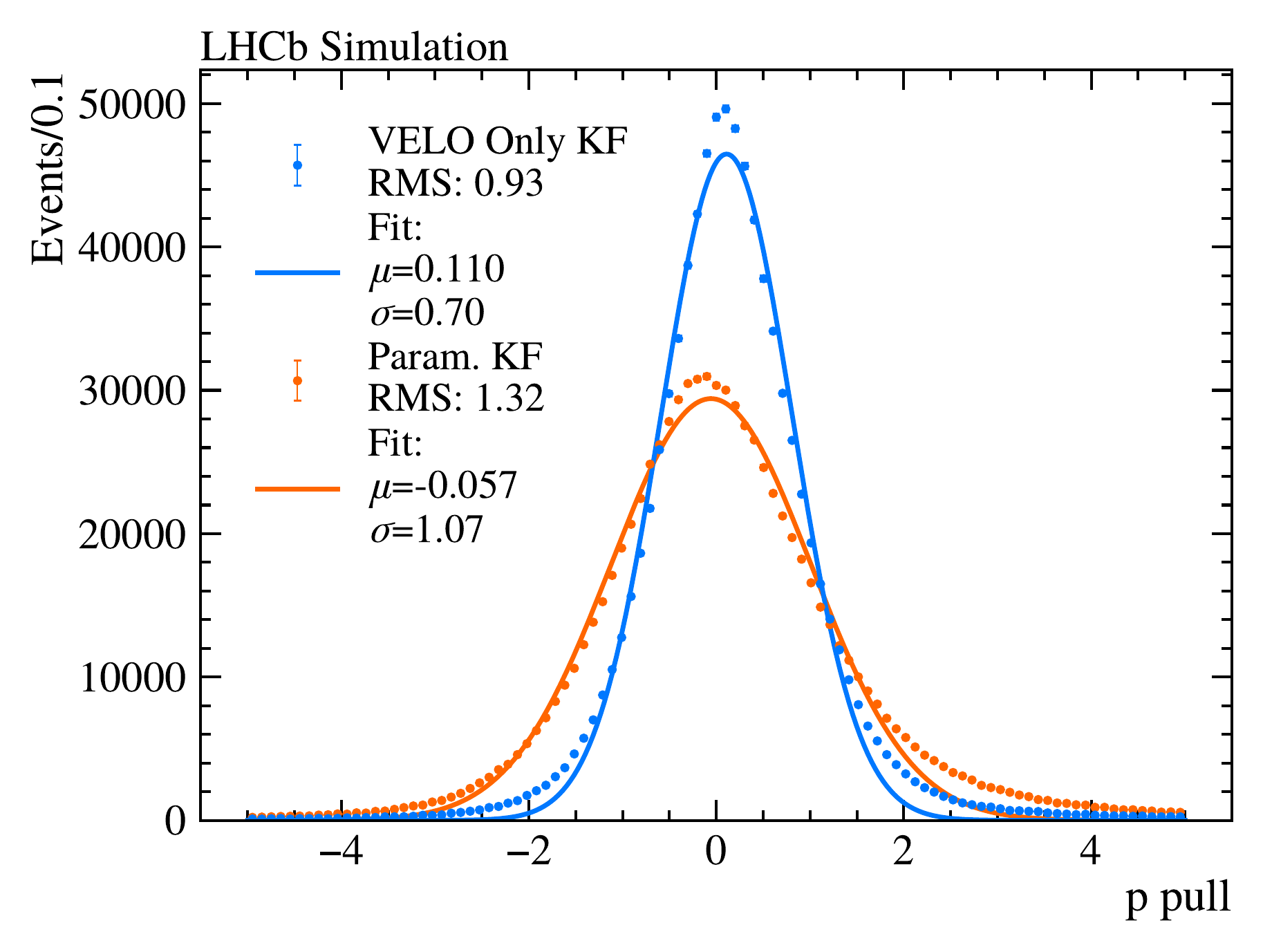}
  \caption{Pull distribution of the momentum estimate.}
  \label{fig:p_pulls}
\end{subfigure}%
\hspace{0.05\textwidth}
\begin{subfigure}[t]{.45\textwidth}
  \centering
  \includegraphics[width=0.95\linewidth]{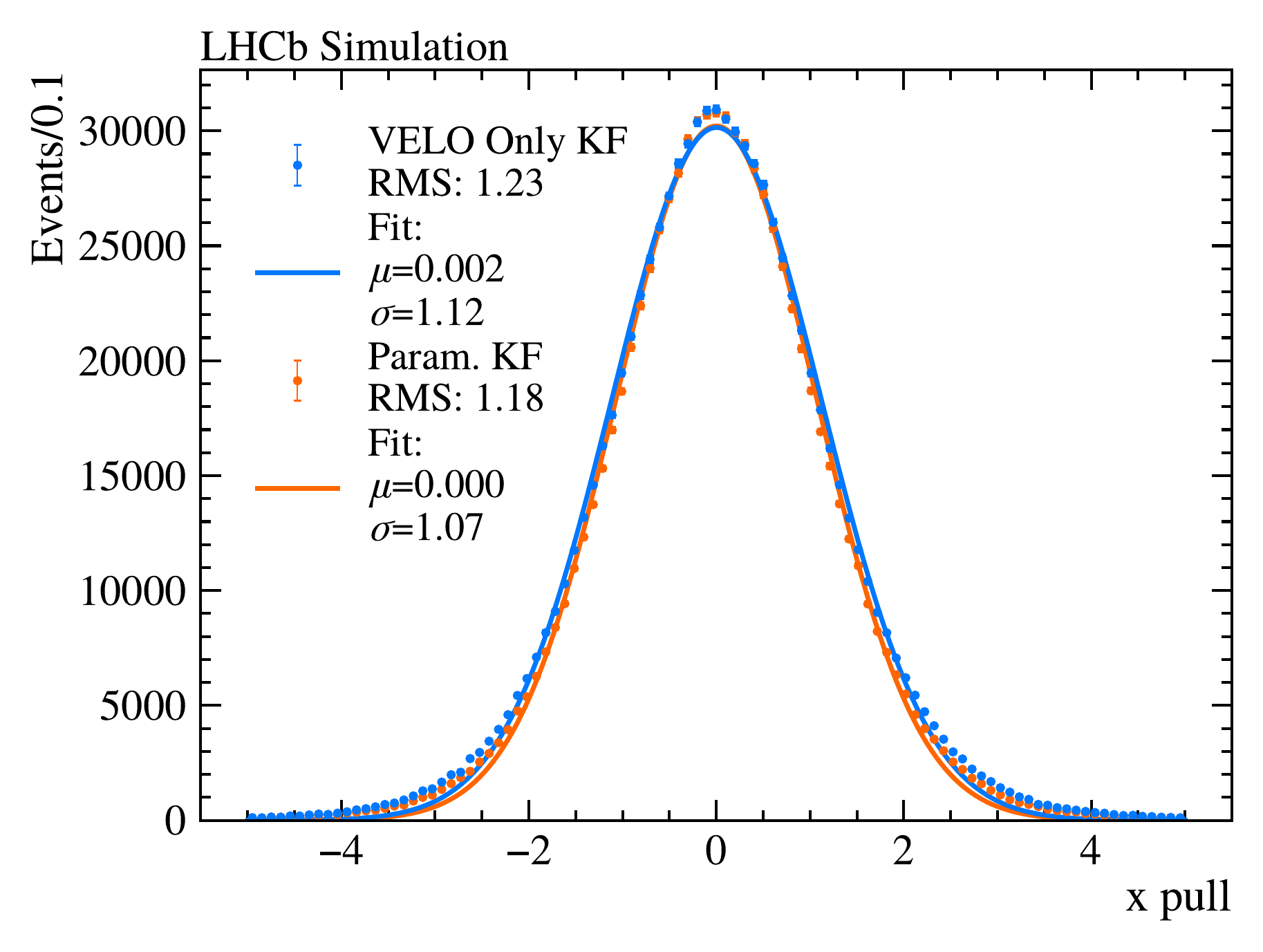}
  \caption{Pull distribution of the $x$ position estimate closest to the beamline.}
  \label{fig:x_pulls}
\end{subfigure} 
\caption{The pulls of (left) the momentum component and (right) the $x$ component of the estimated states. The histograms show the quantities using the same simulated event sample, once using \vokf\ and once using \parkf, but otherwise identical reconstruction sequences.}
\label{fig:pulls}
\end{figure}

\begin{figure}
\centering
\begin{subfigure}[t]{.45\textwidth}
  \centering
  \includegraphics[width=.95\linewidth]{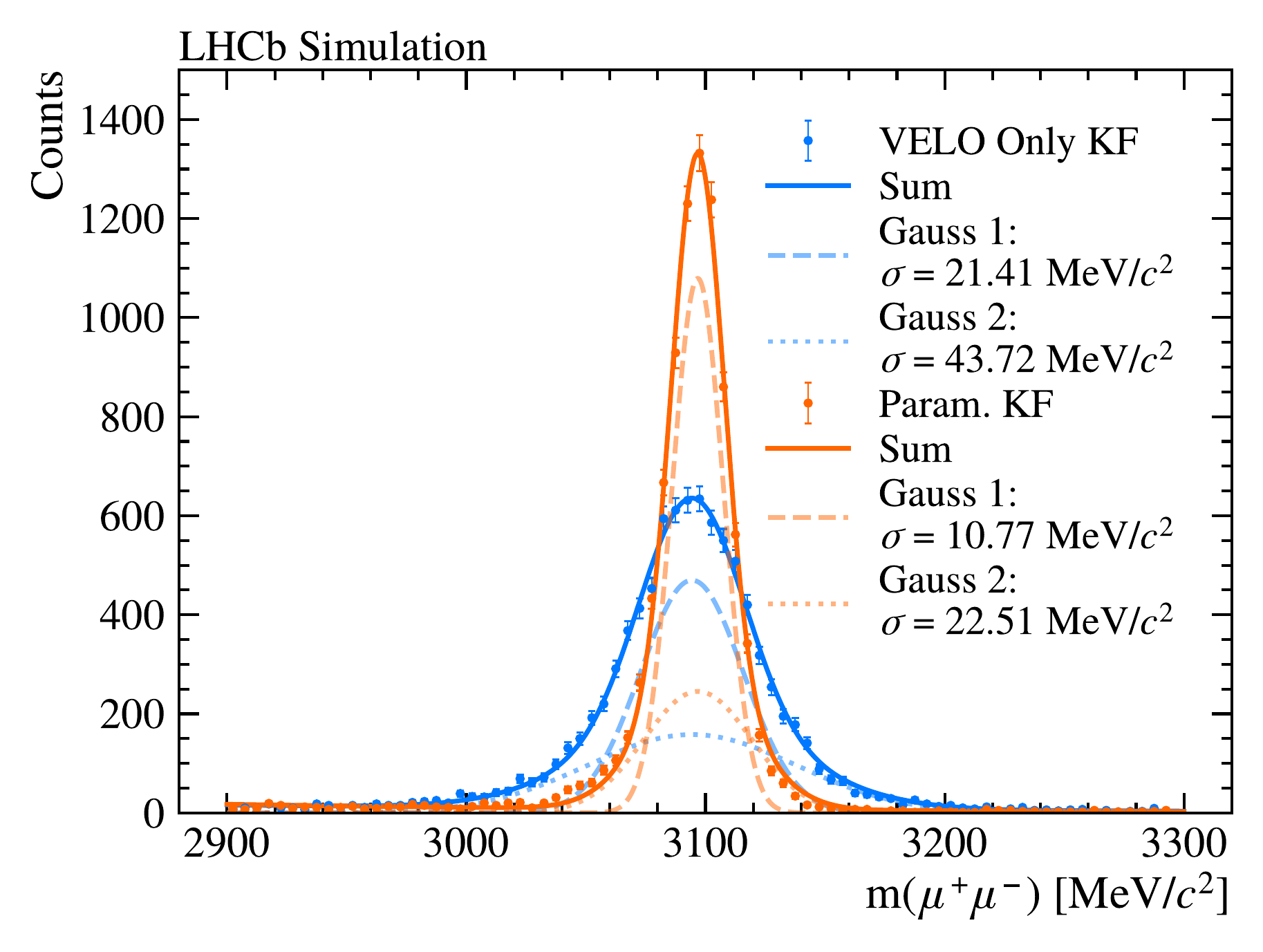}
  \caption{Invariant mass histogram of dimuon-vertices created as part of the `Hlt1DiMuonHighMass' \hltone line with a \jpsi mass peak.}
  \label{fig:jpsi_mass}
\end{subfigure}%
\hspace{0.05\textwidth}
\begin{subfigure}[t]{.45\textwidth}
  \centering
  \includegraphics[width=0.95\linewidth]{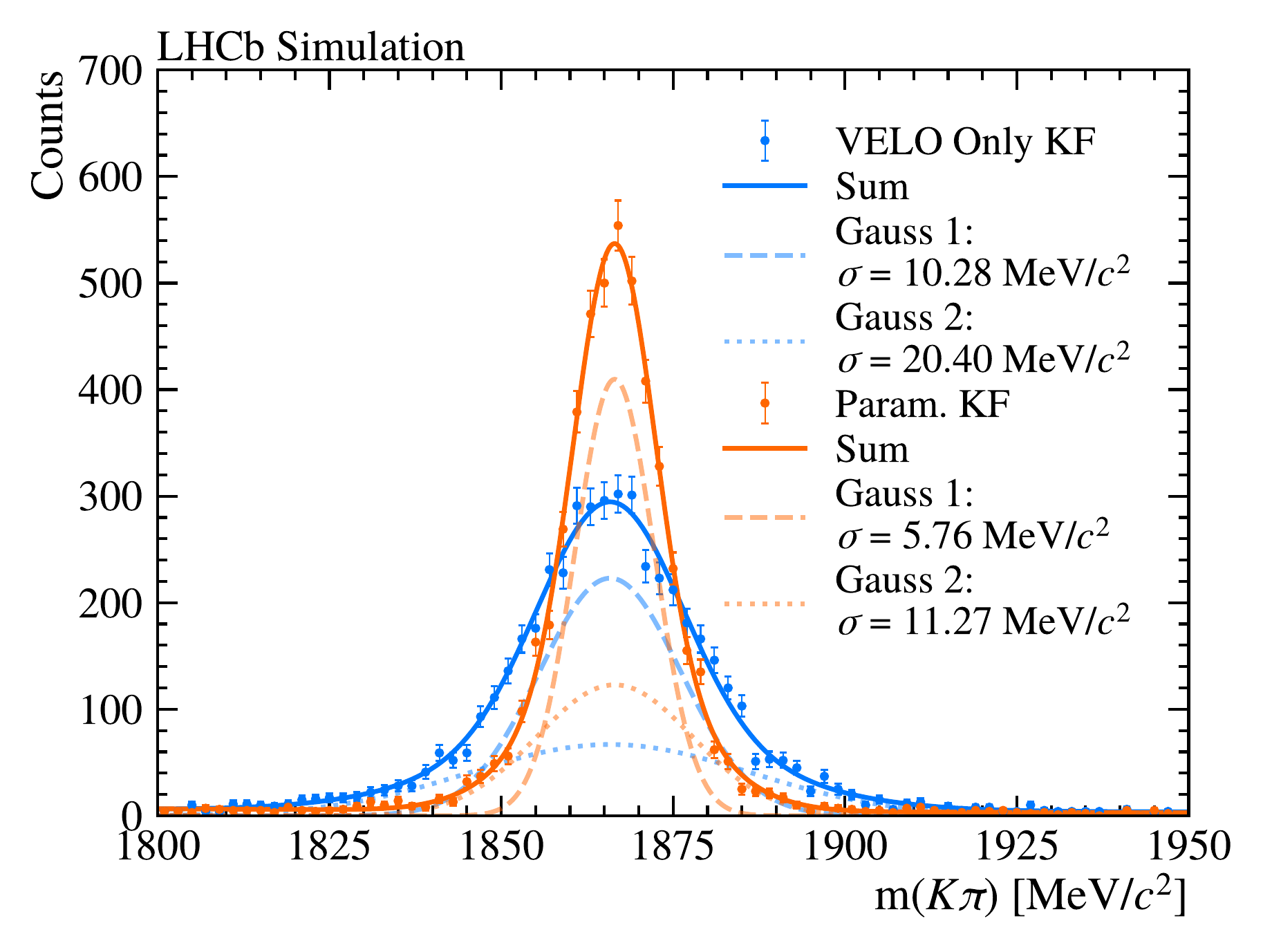}
  \caption{Invariant mass histogram of two-track combinations created as part of the `Hlt1D2KPi' \hltone line with a \Dz mass peak. Note that \hltone does not reconstruct the necessary detector information to distinguish between \pion and \kaon.}
  \label{fig:pion_mass}
\end{subfigure} 
\caption{Two-particle invariant mass histograms taken from running \hltone lines over simulation of signal events. The histograms compare the reconstruction of the same simulated event sample, once using \vokf\ and once using \parkf, but otherwise identical reconstruction sequences. The data was selected using (left) oppositely charged dimuon mass pairs and (right) oppositely charged pairs of kaons and pions.}
\label{fig:mass}
\end{figure}

The state estimates and their covariance are examined for two components in Fig.~\ref{fig:pulls} by calculating the pulls. The pull distribution of the momentum estimate shows a strong improvement, though some non-gaussian behaviour is still apparent. However, covariance of the momentum is not used further in the 
\hltone processing, and the bias and precision is better understood using Fig.~\ref{fig:p_all}. For the spatial component $x$, which is dominated by the VELO component of the fit, the improvements are smaller, as the \vokf\ has a similar approach for the VELO segment of the track fit.
Improvements in momentum resolution and general state resolution are apparent in the improved resolution of mass peaks that are reconstructed in simulated signal samples. Figure \ref{fig:mass} shows invariant mass spectra resulting from dihadron and dimuon combinations of tracks, which include \Dz and \jpsi mass peaks. The widths of the mass peaks are fitted with double Gaussians. Significant improvements close to a factor 2 in the resolution can be seen compared to the \vokf. A similar comparison of the dimuon resolution using \decay{\jpsi}{\mumu} decays with data recorded in 2025 is available in Ref.~\cite{LHCB-FIGURE-2025-020}.

\section{Conclusion}
\label{sec:Conclusion}

This paper presents the implementation and deployment of the \parkf\ in the \lhcb \hltone trigger system. 
The algorithm has been operational since the beginning of the 2025 data-taking period, demonstrating the feasibility of performing full-track Kalman filtering at the 30\mhz \lhc collision rate, using GPU computing.
The core development lies in replacing computationally expensive magnetic field map lookups and Runge-Kutta methods with fast parameterisations, while maintaining the mathematical rigour of the Kalman filter framework. This approach reduces the algorithm to simple functions that are well-suited to GPU architecture limitations, particularly the constraints on fast memory access and the requirement for single-precision arithmetic.

The performance improvements are shown across multiple metrics. Momentum resolution shows strong improvement, with the \parkf\ almost matching the momentum resolution of the full reconstruction in \hlttwo. This is particularly evident in invariant mass distributions where \jpsi and \Dz peak widths are almost halved. Crucially, the \parkf\ demonstrates robustness against detector misalignment effects that degrade the performance of momentum estimates based solely on pattern recognition, making it more reliable for stable physics performance throughout long data-taking periods.
The algorithm also provides a better track-quality metric through improved $\chi^2$ distributions, enabling better separation between genuine particle tracks and combinatorial background. This enhanced track quality directly translates to improved trigger efficiency and a reduced number of background events, while from a computational perspective, the \parkf\ increases the \hltone processing time only by about 2\%. 

The successful deployment of the \parkf\ in \lhcb's production trigger system demonstrates the potential for advanced reconstruction algorithms in real-time high-energy physics computing and is another step towards an offline-quality reconstruction on GPUs. The improved track parameter estimates enable more effective event selection in the trigger system, increasing the efficiency for recording events of physics interest and thereby enhancing the statistical precision of \lhcb's measurements in Run 3 and beyond. 

Further developments include the implementation of outlier removal and the addition of state estimates for a future particle identification using reconstruction in the RICH detectors~\cite{LHCB-FIGURE-2024-027}. The \parkf\ and its future enhancements represent a step towards an offline-quality reconstruction at the full LHC collision frequency, which is envisaged for the next upgrade of the LHCb experiment\cite{LHCb-TDR-023}. 

\section{Acknowledgments}
We thank LHCb’s Real-Time Analysis project for the support, for many useful discussions, and for reviewing an early version of this manuscript. We also thank the LHCb Computing, Simulation and Online projects for producing the simulated samples used to test and benchmark the performance of the parameterised Kalman filter and to evaluate the HLT1 throughput. M. De Cian acknowledges support from STFC, L. H. Uecker acknowledges support from BMFTR.